\documentclass{llncs}
\usepackage[utf8]{inputenc}
\usepackage[T1]{fontenc}
\usepackage{amssymb}
\usepackage{amsmath}
\usepackage{tabularx}
\usepackage{graphicx}
\usepackage{xcolor}
\usepackage{comment}
\usepackage{soul}
\usepackage{hhline}
\usepackage{hyperref}
 
\newcommand{\norm}[1]{\left\lVert#1\right\rVert}

\begin{document}
\title{Node Overlap Removal Algorithms:\\
A Comparative Study}
\titlerunning{Node Overlap Removal Algorithms}
\author{Fati Chen \inst{1, 2}\orcidID{0000-0002-9203-4229} \and
    Laurent Piccinini\inst{2}\and
    Pascal Poncelet\inst{1}\and
    Arnaud Sallaberry \inst{1, 2}\orcidID{0000-0001-7068-176X}}

\authorrunning{F. Chen \textit{et al.}}
\institute{LIRMM - CNRS - Universit\'e de Montpellier, Montpellier, France firstname.lastname@lirmm.fr\and
    Universit\'e Paul-Val\'ery Montpellier 3, Montpellier, France firstname.lastname@univ-montp3.fr}
\maketitle              % typeset the header of the contribution
\begin{abstract}
Many algorithms have been designed to remove node overlapping, and many quality criteria and associated metrics have been proposed to evaluate those algorithms. Unfortunately, a complete comparison of the algorithms based on some metrics that evaluate the quality has never been provided and it is thus difficult for a visualization designer to select the algorithm that best suits his needs. In this paper, we review 21 metrics available in the literature, classify them according to the quality criteria they try to capture, and select a representative one for each class. Based on the selected metrics, we compare 8 node overlap removal algorithms. Our experiment involves 854 synthetic and real-world graphs. 
\end{abstract}

\section{Introduction}

Graph  drawing  algorithms  are  good  at  creating  rich  expressive graph layouts but often consider nodes as points with no dimensions. After changing the size of nodes in the case of annotation or evolving graphs, it causes node overlap which hides information. Post-process algorithms, named \textit{layout adjustment}~\cite{misue1995layout}, have been proposed to remove node overlap.

The objective of these algorithms is, given an initial positioning of the nodes and a size for each one, to provide a new embedding so that there are no overlapping nodes any more. A classical zoom-in function maintaining the sizes of the nodes (i.e. uniform scaling) provides such an embedding, but it expands the visualisation, resulting in large areas without any objects. Therefore, a node overlap removal algorithm must take into account the area of the drawing, and try to minimise it. Positioning the nodes evenly on a grid meets this objective but will result in the loss of the user's mental picture of the original embedding. Thus, it is also important to minimise the change on the layout.

Since a preliminary work in 1995~\cite{misue1995layout}, many algorithms have been designed to reach these purposes, and many quality criteria have been proposed to evaluate them. Unfortunately, a complete comparison of the algorithms based on the different criteria has never been provided and it is thus difficult for a visualisation designer to select the one that best suits his needs.

In this paper, our contribution comes in two forms: (1) We propose a classification of 21 quality metrics, grouping them according to the quality criterion they try to capture. We also discuss their relevance and we select a representative one for each class. (2) We compare state-of-the-art node overlapping approaches in regards to the previously  selected metrics. This experiment involves 854 graphs, including synthetic ones (random, tree, scale-free, small-world) and real world ones.

The paper is organised as follows: after a brief reminder in Sec.\ref{sec:preliminaries} of the definitions and the notations used in this paper, we present and discuss the quality criteria and the metrics in Sec. \ref{sec:criterias}. Then we compare the algorithms in Sec. \ref{sec:comparison}. Finally we conclude in Sec. \ref{sec:conclusion}.

\section{Preliminaries}\label{sec:preliminaries}

In this paper, we use the following definitions and notations. 

$G = (V, E)$ denotes a graph where $V$ is the set of nodes and $E$ the set of edges. The number of nodes $|V|$ is denoted by $n$ and the number of edges $|E|$ by $m$. We consider each node as a rectangle. Thus, for a node $v\in V$, its width and its height are denoted by the couple $(w_v, h_v)$ which is not impacted by the layout adjustment.

The initial embedding is defined as an injection $\mathcal{E}_G : V \rightarrow \mathbb{R}^2$ such that $\forall v\in V$, $\mathcal{E}_G(v) = (x_v, y_v)$ where $(x_v, y_v)$ are the coordinates of the center of the node $v$. The overlapping-free embedding is denoted by $\mathcal{E}'_G$. 
To simplify notations, we denote $v = (x_v, y_v)$ instead of $\mathcal{E}_G(v)$, and $v' = (x'_v, y'_v)$ instead of $\mathcal{E}'_G(v)$. Remark that two nodes $(u, v) \in V^2$ are overlapping when :
$$
    |x_v - x_u|  < \frac{w_v + w_u}{2} \quad\text{and}\quad
    |y_v - y_u|  < \frac{h_v + h_u}{2}
$$

The bounding box $Bb$ of an embedding $\mathcal{E}_G$ is defined as the smallest rectangle containing all the nodes of $G$; $w_{bb}$ (resp. $h_{bb}$) denotes the width (resp. the height) of the initial embedding, $w'_{bb}$ (resp. $h'_{bb}$) denotes the width (resp. the height) of the overlapping-free one. They are determined as follows:
\begin{align}
    w_{bb} & = \left|\max_{v \in V}\left( x_v + \frac{w_v}{2}\right) - \min_{u \in V}\left( x_u - \frac{w_u}{2}\right)\right| \\
    h_{bb} & = \left|\max_{v \in V}\left( y_v + \frac{h_v}{2}\right) - \min_{u \in V}\left( y_u - \frac{h_u}{2}\right)\right|
\end{align}
The position of the center of the bounding box is denoted by $c_{bb}=(x_{bb},y_{bb})$ in the initial embedding, and $c'_{bb}=(x'_{bb},y'_{bb})$ in the overlapping-free embedding.

The convex hull of an embedding $\mathcal{E}_G$ is defined as the smallest convex region containing all the nodes of $G$. Note that it is computed by using the 4 corners of the nodes, and not only their center, in a way that the rectangles representing the nodes are fully included into it. In the following, $Ch$ denotes the convex hull of the original embedding, $Ch'$ the convex hull of the free-overlapping one, $c_{ch}$ the center of mass of $Ch$, $c'_{ch}$ the center of mass of $Ch'$. 

\section{Quality Criteria}\label{sec:criterias}

Many criteria have been proposed in the literature to evaluate the quality of the embeddings resulting from adjustment algorithms. Unfortunately, the experiments provided by the authors of the different approaches are not always based on the same metrics. With a view to provide a uniform protocol of experiment and a complete comparison of the algorithms, we need to review the quality criteria and the metrics used to evaluate them. We also need to select a representative metric for each criterion.

We identified 5 classes of metrics (\textit{Orthogonal Ordering preservation}, \textit{Spread minimisation}, \textit{Global Shape preservation}, \textit{Node Movement minimisation} and \textit{Edge Length preservation}), each of them depicting a quality criterion. Table~\ref{tab:criterias} shows the metrics assigned into the classes. The formulas are given in the discussion below.

\begin{table}[ht!]
    \caption{List of metrics classified by the quality criteria they try to capture: metrics selected for the comparison appear in bold italics. The \textit{Abbreviations} are based on some initials of the names. For example, $sp\_bb\_a$ means that the metric is in the class \textit{Spread minimisation}, it uses the embedding \textit{Bounding Box} to quantify the \textit{Area} spreading. The \textit{Range} column contains the set of values that the metric can take. The \textit{Target} column refers to the target value to meet the corresponding criterion.}
    \label{tab:criterias}
    \begin{tabular}{|l l|c|c|}
        \hline
        {\bf Abbreviation}    & {\bf Name}                                                       & {\bf Range} & {\bf Target} \\\hline
        \hline         & \textbf{Orthogonal Ordering preservation}                                     & &                              \\\hline
        $oo\_o$        & Original \cite{misue1995layout}                                  & $\{0,1\}$      & 1             \\\hline
        $oo\_kt$       & Kendall's Tau Distance \cite{huang2007force}                     & $[0,1]$        & 0             \\\hline
        $oo\_ni$       & Number of Inversions \cite{strobelt2010roller}                   & $[0, n(n -1)]$ & 0             \\\hline
        $oo\_nni$      & \textbf{\textit{Normalised Number of Inversions}}                   & $[0, 1]$       & 0             \\\hline
        \hline         & \textbf{Spread minimisation}                                                  &                &               \\\hline
        $sp\_bb\_l1ml$ & Bounding Box L1 Metric Length \cite{li2005spring}                & $[1, +\infty[$ & 1             \\\hline
        $sp\_bb\_a $   & Bounding Box Area \cite{misue1995layout}                         & $[1, +\infty[$ & 1             \\\hline
        $sp\_bb\_na$   & Bounding Box Normalised Area \cite{huang2007force}               & $[0, 1[$       & 0             \\\hline
        $sp\_ch\_a$    & \textbf{\textit{Convex Hull Area}} \cite{strobelt2010roller}                       & $[1, +\infty[$ & 1             \\\hline
        \hline         & \textbf{Global Shape preservation}                               &                &               \\\hline %
        $gs\_bb\_ar$   & Bounding Box Aspect Ratio \cite{li2005spring}                    & $]0, +\infty[$ & 1             \\\hline
        $gs\_bb\_iar$  & \textbf{\textit{Bounding Box Improved Aspect Ratio}}                 & $[1, +\infty[$ & 1             \\\hline
        $gs\_ch\_sd$   & Convex Hull Standard Deviation \cite{strobelt2010roller}         & $[0, +\infty[$ & 0             \\\hline
        \hline         & \textbf{Node Movement minimization}                                           &                &               \\\hline %
        $nm\_mn$       & Moved Nodes \cite{huang2007force}                                & $[0, 1]$       & 0             \\\hline
        $nm\_dm\_me$   & Distance Moved Mean Euclidean \cite{strobelt2010roller}          & $[0, +\infty[$ & 0             \\\hline
        $nm\_dm\_ne$   & Distance Moved Normalized Euclidean \cite{lyons1998cluster}      & $[0,1]$        & 0             \\\hline
        $nm\_dm\_h$    & Distance Moved Hamiltonian \cite{huang2003force,huang2007force}  & $[0, +\infty[$ & 0             \\\hline
        $nm\_dm\_se$   & Distance Moved Squared Euclidean \cite{marriott2003removing}     & $[0, +\infty[$ & 0             \\\hline
        $nm\_dm\_imse$ & \textbf{\textit{Distance Moved Improved Mean}}      &  &            \\
        &  \textbf{\textit{Squared Euclidean}}      & $[0, +\infty]$ & 0             \\\hline
        $nm\_d$        & Displacement \cite{gansner2010efficient}                         & $]0, +\infty[$ & 0             \\\hline
        $nm\_knn$      & K-Nearest Neighbours \cite{nachmanson2016node}                    & $[0, +\infty[$ & 0             \\\hline
        \hline         & \textbf{Edge Length preservation}                                              &                &               \\\hline %
        $el\_r$        & Ratio \cite{li2005spring}                                        & $[1, +\infty[$ & 1             \\\hline
        $el\_rsdd$     & \textbf{\textit{Relative Standard Deviation Delaunay}} \cite{gansner2010efficient} & $[0,+\infty]$  & 0             \\\hline
    \end{tabular}
\end{table}

The following subsections contain the metrics of a specific class. In each of them, we select one representative metric, based on the corresponding quality criterion and the properties that the metrics aim at capturing. Our discussion also sometimes involves the coefficient of correlation of two metrics run following the protocol described in the comparison section, Sec.~\ref{sec:comparison}. 

% Orthogonal ordering %
%=====================%
\subsection{Orthogonal Ordering Preservation}

The orthogonal ordering class groups the metrics which try to quantify how much an adjustment algorithm preserves the initial orthogonal ordering. We recall that the orthogonal ordering is respected when all nodes satisfy the following conditions:
\[
\begin{cases}
x_u < x_v \Leftrightarrow x'_u < x'_v \\
y_u < y_v \Leftrightarrow y'_u < y'_v \\
x_u = x_v \Leftrightarrow x'_u = x'_v \\
y_u = y_v \Leftrightarrow y'_u = y'_v
\end{cases}
\]
    
The first metric of this class, $oo\_o$ \cite{misue1995layout}, is equal to $1$ if the overlapping-free graph embedding preserves the initial orthogonal ordering, $0$ otherwise. Also, if only one couple of nodes does not satisfy those conditions, the value of $oo\_o$ is the same as when many ones do not satisfy it.

To overcome this issue, Huang \textit{et al.}~\cite{huang2007force} proposed a metric based on the \emph{Kendall’s Tau distance}. For each couple of nodes, they first compute an inversion number $\textit{inv}(u, v)$ corresponding to $0$ if the orthogonal ordering is preserved between them, $1$ otherwise. The metric is then defined as the normalised sum of the inversion numbers:
\[
    \text{oo\_kt} = \frac{%
        \displaystyle\sum_{u\not= v}{\textit{inv}(u, v)}
    }{n(n-1)}
\]

Strobelt \textit{et al.}~\cite{strobelt2010roller} introduced the number of inversions: 
\begin{align*}
    \text{oo\_ni} =  &\sum_{\overset{(u, v) \in V^2}{x_u > x_v}}\begin{cases}
        1 & \text{if }x'_u < x'_v \\
        0 & \text{otherwise}
    \end{cases} \\
           + &\sum_{\overset{(u, v) \in V^2}{x_u > x_v}}\begin{cases}
        1 & \text{if }y'_u < y'_v \\
        0 & \text{otherwise}
    \end{cases}
\end{align*}

This metric has the drawback of providing non-normalized values. However, it holds the benefit of penalizing inversions occurring on each axis independently ($x-$ and $y-axis$), instead of penalizing in the same manner an inversion occurring in only one axis and an inversion occurring in the two axes. Thus, in our study, we combine the two metrics by using a normalised version of the latter:
    \[
        \text{oo\_nni} = \frac{oo\_ni}{n(n-1)}
    \]

%* Spread %
%*========%
\subsection{Spread Minimisation}

A classical zoom-in function maintaining the sizes of the nodes (i.e. uniform scaling) provides an overlapping-free embedding, but it expands the visualisation, resulting in large areas without any objects. To avoid this issue, quality metrics have been introduced to quantify embedding spreading. Their purpose is to favour algorithms inducing low spreading.
 
The L1 metric length~\cite{li2005spring} is the ratio:

\[
    \text{sp\_bb\_l1ml} = \frac{\max(w'_{bb}, h'_{bb})}{\max(w_{bb}, h_{bb})}
\]

The drawback of this technique is to consider only one dimension of the embedding, width or height. For instance, considering an example where $w_{bb}=4$, $h_{bb}=2$, $w'_{bb}=4$, $h'_{bb}=4$, the value of the L1 metric length is $1$ (which is the target value), whereas the area of the overlapping-free embedding is twice as large as in the initial embedding. The ratio between the bounding box areas of the two embeddings~\cite{misue1995layout} overcomes this issue: 

\[
    \text{sp\_bb\_a} = \frac{w'_{bb} \times h'_{bb}}{w_{bb} \times h_{bb}}
\]

While the result gives an unbounded value greater than $1$, Huang \textit{et al.}~\cite{huang2007force} proposes a normalised version producing values in the interval $[0, 1[$:

\[
    \text{sp\_bb\_na} = 1 - \frac{w_{bb} \times h_{bb}}{w'_{bb} \times h'_{bb}}
\]

Unfortunately, this criterion is poorly intuitive and it is hard to figure out what the values represent.

In our comparison, we selected another version of the ratio of areas involving convex hulls~\cite{strobelt2010roller}, as it better captures the concrete area of the drawing:
\[
    \text{sp\_ch\_a} = \frac{area(Ch')}{area(Ch)}
\]

% GLOBAL SHAPE %
%==============%
\subsection{Global Shape Preservation}

This class contains metrics that try to capture the ability of the algorithms to preserve the global shape of the initial embedding. The first one was proposed by Li \textit{et al.} ~\cite{li2005spring}: 

\[
    \text{gs\_bb\_ar} = \begin{cases}
        \text{if } w'_{bb} > h'_{bb}
         & \dfrac{w'_{bb} \times h_{bb}}{h'_{bb} \times w_{bb}} \\
        \text{otherwise}
         & \dfrac{h'_{bb} \times w_{bb}}{w'_{bb} \times h_{bb}}
    \end{cases}
\]

The underlying idea is to capture the variation of the aspect ratio ($w_{bb}/h_{bb}$) between the initial and the overlapping-free embedding. For instance, let us consider an example where $w_{bb}=3$, $h_{bb}=2$, $w'_{bb}=6$, $h'_{bb}=4$. In this case, the overlapping-free embedding is twice as large as the initial one but the aspect ratio remains the same $3/2$. The $gs\_bb\_ar$ is 1, which is the target value. Now let us consider another example where $w_{bb}=3$, $h_{bb}=2$, $w'_{bb}=4$, $h'_{bb}=6$. In this case, the initial aspect ratio is $3/2$ whereas the overlapping-free one is $2/3$. The $gs\_bb\_ar$ is now $2.25$, which is not the target value; it reveals a distortion of the initial embedding during the overlap removal process. The main drawback of this metric is that it can reach values in the interval $]0, +\infty[$ while the target value is $1$. Thus, it is hard to decide, for instance, which algorithm is the best between two of them if the first one obtains a score of $0.67$ and the second one a score of $4.56$. To overcome this issue, we propose to refine it as follows: 

\[
    \text{gs\_bb\_iar} = \max{\left(\frac{w'_{bb} \times h_{bb}}
        {h'_{bb} \times w_{bb}}
        , \frac{h'_{bb} \times w_{bb}}
        {w'_{bb} \times h_{bb}}
        \right)}
\]

In this case, the target value is $1$ and the metric cannot reach values below it. This criterion is the one we selected for our study.

An alternative to this approach based on the convex hull has been proposed by Strobelt \textit{et al.}~\cite{strobelt2010roller}. The idea is to evaluate the distortion of the convex hull by comparing, between both embeddings, the distances of convex hull points to their center. Let $\ell_\theta$ (resp. $\ell'_\theta$) be the euclidean distance between the center of mass $c_{ch}$ (resp. $c'_{ch}$) of the convex hull $Ch$ (resp. $Ch'$) and the intersection of the convex hull with the line going through $c_{ch}$ (resp. $c'_{ch}$) and with an angle $\theta$ ($\theta$ varying from $0^{\circ}$ to $350^{\circ}$ in $10^{\circ}$ steps). Then, the difference is defined as the ratio $d_\theta = \ell'_\theta/\ell_\theta$. The metric is the standard deviation of the 36 measures of $d_\theta$:
\[
    \text{gs\_ch\_sd} = \sqrt{\frac{1}{36}\sum_{\underset{k=0,\cdots,35}{\theta = 10k}}\!\!\!\!(d_\theta - \overline{d})^2}
\]
\[
\text{ \ where } 
    \overline{d}=\frac{1}{36}\sum_{\underset{k=0,\cdots,35}{\theta = 10k}}\!\!\!\!d_\theta
\text{ is the mean value}
\]

\begin{comment}
The metric is the standard deviation of the difference values for angles varying from $0^{\circ}$ to $350^{\circ}$ in $10^{\circ}$ steps:

\[
    \text{gs\_ch\_sd} = \sqrt{\frac{1}{36}\sum^{35}_{a = 0}(d_{a \cdot 10^\circ} - \overline{d})}
\]

where 
\[
    \overline{d}=\frac{1}{36}\sum^{35}_{a = 0}d_{a \cdot 10^\circ}
\]
\end{comment}

Based on the experiments presented below in Sec.~\ref{sec:comparison}, we  observed that $gs\_bb\_iar$ and $gs\_ch\_sd$ have a correlation coefficient of $0.77$, showing that they both tend to capture similar aspects of the adjustment process. We selected the former for its simplicity and its ease of interpretation.

% Node Movement %
%================%
\subsection{Node Movement Minimisation} 

This class contains the metrics quantifying the changes in node positions after running an adjustment algorithm. The underlying intuition is that an algorithm involving high node movements will provide an overlapping-free configuration different from the original one, and thus may result in a substantial loss of the mental model.

The simplest metric of this class was presented by Huang \textit{et al}.~\cite{huang2007force}: 
\[
     \text{nm\_mn} = \frac{nb}{n}
\]
Here, $nb$ represents the number of nodes which have moved between the initial and the overlapping-free embedding. The main drawback of this approach is that a node overlap removal algorithm may induce very small changes in most nodes, which does not affect the mental model preservation, while inducing a very bad result. To tackle this problem and add more granularity over the evaluation of node movements, a series of metrics have been proposed, based on the same underlying quality function:
\[
    \text{nm\_dm} = f(n) \times \sum_{v \in V}{\text{dist}(v, v')}
\]
where $f$ is a normalising function of $n=|V|$ and $dist$ is a distance between $v$ and $v'$. Table~\ref{tab:fdist} sums up the ones used in the literature.
\begin{table}[ht]
        \caption{Functions used to tune the distances moved metric.}    \label{tab:fdist}
    \begin{tabular}{|>{\centering\arraybackslash}p{4 cm}||c|c|c|}
        \hhline{-||-|-|-|}
        $\text{dist}(v, v')\quad$\textbackslash $\quad f(n)$ & $1$ & {\bf $1/n$} & {\bf $\frac{1}{k\sqrt{2}\times n}$} \\ \hhline{=::=:=:=}
        $\norm{v' - v}$ & & $nm\_dm\_me$ \cite{strobelt2010roller} & $nm\_dm\_ne$ \cite{lyons1998cluster} \\ \hhline{-||-|-|-|}
        $\norm{v' - v}^2$ & $nm\_dm\_se$ \cite{marriott2003removing} & $nm\_dm\_imse$ & \\ \hhline{-||-|-|-|}
        $|x'_v - x_v| + |y'_v - y_v|$ & $nm\_dm\_h$ \cite{huang2003force} & & \\ \hhline{-||-|-|-|}
    \end{tabular}
\end{table}

The function $f$ comes in three different forms. Marriott \textit{et al.}~\cite{marriott2003removing} and Huang \textit{et al.}~\cite{huang2003force} do not include any $f$, which is similar to having $f(n)=1$. The drawback is that the resulting value highly depends on the number of nodes in the graph. That is why Strobelt \textit{et al.} \cite{strobelt2010roller} proposed to use the mean of the distances, which corresponds to $f(n)=1/n$. Finally, Lyons \textit{et al.}~\cite{lyons1998cluster} proposed $f(n)=1/(k\sqrt{2}\times n)$, where $k$ is the maximum between $w'_{bb}$ and $h'_{bb}$. In this case, $k\sqrt{2}$ is the diagonal of a square containing the embedding, thus a maximum distance available for a node. Thus, this $f$ function normalises the values of the metric. Unfortunately, this normalisation induces very small values that are hard to interpret. That is why we preferred using $f(n)=1/n$ for our study.

Three $dist$ functions have been proposed in the literature. The most intuitive one is the Euclidean distance $\norm{v' - v}$~\cite{strobelt2010roller,lyons1998cluster}. The squared Euclidean distance $\norm{v' - v}^2$~\cite{marriott2003removing} avoids the square root computation and discriminates high changes better. It is the one we selected for our study. The Manhattan distance $|x'_v - x_v| + |y'_v - y_v|$ has also been used~\cite{huang2003force}, but it is less intuitive and has close results ($nm\_dm\_se$ and $nm\_dm\_h$ have a correlation coefficient of 0.9).

Let us consider an adjustment algorithm that pushes nodes on the x-axis. The preservation of the global shape is not optimal but the preservation of the configuration should reach a good score, as a node on right-top in the initial embedding would remain on right-top in the overlapping-free embedding. In order to better capture the relative movement of a node between the two embeddings, a $shift$ function can be applied to align the center of the initial bounding box with the center of the final one, and a $scale$ function to align the size of the initial bounding box to the size of the final one: 
\begin{align*}
     shift(v) =& \left( x_v + x'_{bb} - x_{bb}, y_v + y'_{bb} - y_{bb} \right) \\  
     scale(v) =& (x_v \times \frac{w'_{bb}}{w_{bb}}, y_v \times \frac{h'_{bb}}{h_{bb}})
\end{align*}

Considering this, we selected the following node movement metric:
\[
    \text{nm\_dm\_imse} = \frac{1}{n} \times \sum_{v \in V}{\norm{v' - scale(shift(v))}^2}
\]
$nm\_d$~\cite{gansner2010efficient} (the complete formula is available in the paper) is also based on the idea that the metric should be based on modified initial positions to better capture the relative movement of the nodes between the two embeddings. Besides including the $shift$ and the $scale$ functions, it also rotates the initial embedding with an angle $\theta$ that minimizes the distances between the nodes of the initial embedding and the ones of the overlapping-free embedding:
\[
    rotation(v) = (x_v\cos{\theta} - y_v\sin{\theta}, x_v\sin\theta  + y_v\cos\theta )
\]

We have not included the rotation in our experiment as we consider that it can induce a loss of the mental model (think about the recognition of a map turned inside down).

An alternative to quantify how much an overlapping-free configuration may result in a substantial loss of the mental model is to look at the neighbourhoods at the nodes and compare them before and after the adjustment. Based on a $KNN$ approach, Nachmanson \textit{et al.}~\cite{nachmanson2016node} proposed the following metric:
\[
    nm\_knn(k) = \sum_{v \in V} \left(k - |N_k(v) \cap N_k(v')|\right)^2
\]
where $N_k(v)$ (resp. $N_k(v')$) denotes the $k$ nearest neighbours of $v$ (resp. $v'$), in terms of Euclidean distance, in the initial (resp. overlapping-free) embedding. We did not select this metric because, unlike the other metrics of the class, it requires to fix a parameter ($k$).

% Edge Based %
%============%
\subsection{Edge Length Preservation} 

This class contains the two metrics based on edge lengths. The set of edges can be $E$ or can be another set derived from the graph.

Standard force-based layout algorithms tend to produce uniform lengths of edges. Indeed, the first metric of this class captures whether the edge lengths of a graph remain uniform or not after applying an adjustment algorithm~\cite{li2005spring}:
\[
    el\_r = \frac{\max_{(u, v) \in E^2} \norm{u' - v'}}
    {\min_{(u, v) \in E^2}\norm{u' - v'}}
\]
While many layout algorithms are not designed to produce uniform edge lengths, we did not select this approach, which is not related to the mental model preservation for these algorithms. We preferred the next one, based on a Delaunay triangulation. 

Let $E_{dt}$ be the set of edges of a Delaunay triangulation performed on the nodes of the initial embedding. The second metric of this class, $el\_rsdd$, is based on computing the coefficient of variation, also known as the relative standard deviation, of the edge lengths ratio as follows~\cite{gansner2010efficient}:
\[
    r_{uv} = \frac{||u' - v'||}{||u - v||}, \quad (u, v) \in E_{dt}^2
\]
\[
    \overline{r} = \frac{1}{|E_{dt}|}\sum_{(u, v) \in E_{dt}^2}{r_{uv}}
 \]
 \[
    el\_rsdd = \frac{\sqrt{
            \frac{1}{|E_{dt}|}\sum_{(u, v) \in E_{dt}^2}{(r_{uv} - \overline{r})^2}}
    }{\overline{r}}
\]
\section{Algorithms Comparison}\label{sec:comparison}

In this section, we compare 8 algorithms of the literature in terms of quality and running time: uniform \textit{Scaling}, \textit{PFS}~\cite{misue1995layout}, \textit{PFS'}~\cite{hayashi1998layout}, \textit{FTA}~\cite{huang2007force}, \textit{VPSC}~\cite{dwyer2005fast}, \textit{PRISM}~\cite{gansner2010efficient}, \textit{RWordle-L}~\cite{strobelt2010roller}, and  \textit{GTREE}~\cite{nachmanson2016node}. The quality of an overlapping-free embedding is evaluated with the metrics identified in the last section, by following a 3 steps procedure:
\textbf{Step 1: Datasets} We generate 840 synthetic graphs containing 10 to 1,000 nodes. These graphs are provided by 4 generation models available on the OGDF library~\cite{ChimaniGJKKM13}: random graphs~\cite{erdos_1959}, random trees, small world graphs~\cite{watts_1998}, and scale-free graphs~\cite{barabasi_1999}. We also use 14 real-world graphs selected from the $\text{Graphviz}$ test suite\footnote{\url{https://gitlab.com/graphviz/graphviz/blob/master/rtest/graphs/}(accessed: 2019-07)}~\cite{gansner2000graphviz}, previously used by the authors of \textit{PRISM}~\cite{gansner2010efficient} and \textit{GTREE}~\cite{nachmanson2016node}.
\textbf{Step 2: Overlapping-free embedding computation} Synthetic graphs resulting from the first step are initially positioned by the $FM^3$ layout algorithm~\cite{hachul2005fmmm}. Then, we apply the 8 node overlap removal algorithms, thus providing a set of 6,720 overlapping-free graph embeddings. \texttt{Graphviz} test suite graphs are initially positioned by the $SFDP$ layout algorithm~\cite{hu2005efficient} to follow the same baseline embedding as Gansner \textit{et al.}~\cite{gansner2010efficient}. We then apply the 8 node overlap removal algorithms thus providing 112 overlapping-free graph embeddings.
\textbf{Step 3: Metrics computation}
We finally compute the values of the 5 selected metrics on the 6.832 overlapping-free synthetic and real-world graph embeddings. We also measure the computation time of the algorithms.

\subsection{Quality}

\begin{figure}[ht!]
    \centering
    \includegraphics[width=0.9\linewidth]{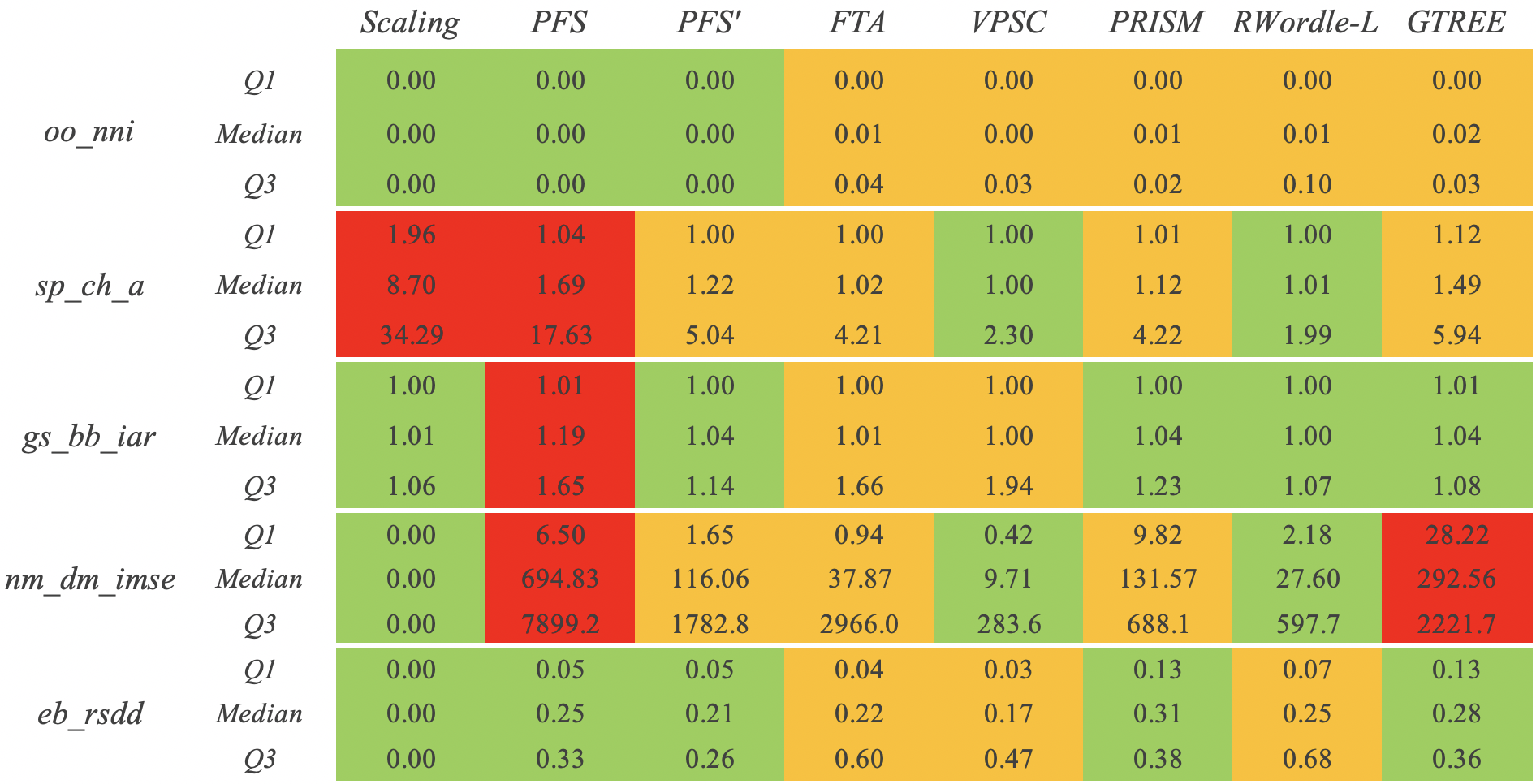}
    \caption{Aggregated values of the selected metrics among the synthetic graphs: first quartile, median and third quartile.}
    \label{fig:res_synt_metrics}
\end{figure}

\begin{figure}[ht!]
    \centering
    \includegraphics[width=0.9\linewidth]{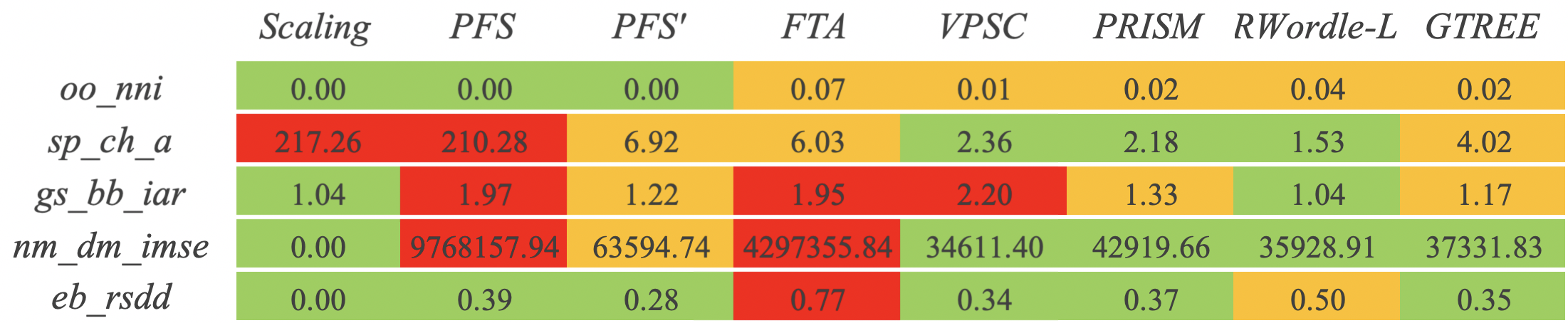}
    \caption{Mean values of the selected metrics among the real-world graphs.}
    \label{fig:res_rw_metrics}
\end{figure}

Fig.~\ref{fig:res_synt_metrics} and~\ref{fig:res_rw_metrics} show the aggregated metrics values on the synthetic and real-world datasets. Unsurprisingly, \textit{Scaling}, \textit{PFS} and \textit{PFS'} obtain the best scores at $oo\_nni$ as it is proved that they maintain the original orthogonal ordering. Though, all the algorithms tested got good results for this criterion. 

\textit{Scaling} highly increases the size of the embedding, which induces a bad score for $sp\_ch\_a$. $PFS$ also obtains a bad score for this criterion. \textit{VPSC} and \textit{RWordle-L} produce the most compact embeddings, while the other algorithms give intermediary results.

\textit{Scaling} preserves the initial global shape\footnote{The global shape preservation score for \textit{Scaling} is not $1$ because of the size of the nodes that remains the same between the initial and the overlapping-free embeddings.} ($gs\_bb\_iar$ score). \textit{PFS} is the worst algorithm on this criterion. The other algorithms obtained good median scores on synthetic graphs, but the third quartile scores show that \textit{FTA} and \textit{VPSC} can produce a certain amount of distorted embeddings. This is confirmed by the tests on real-world graphs, where they obtain worse results.

\textit{Scaling} obtains the best results for the node movement minimisation criterion, followed by \textit{VPSC} and \textit{RWordle-L}. \textit{FTA} also obtained a good median score on synthetic graphs, but its third quartile value shows that it can generate a certain amount of embeddings with high changes, as also illustrated by the bad score obtained on the real-world graphs. \textit{PFS'} and \textit{PRISM} obtained intermediary results. Finally, \textit{GTREE} had bad results on the synthetic graphs, while it obtained pretty good ones on the real-world graphs.

\textit{Scaling} preserves relative edge lengths. All the other criteria obtained comparable median score between 0.17 and 0.31. However, the third quartile on the synthetic graphs shows that \textit{FTA}, \textit{VPSC} and \textit{RWordle-L} generate a certain amount of embeddings with high variations. This observation is confirmed by the results on the real-world graphs for \textit{FTA} and \textit{RWordle-L}.

\subsection{Computation Time}

\begin{figure}[ht!]
    \centering
    \includegraphics[width=0.9\linewidth]{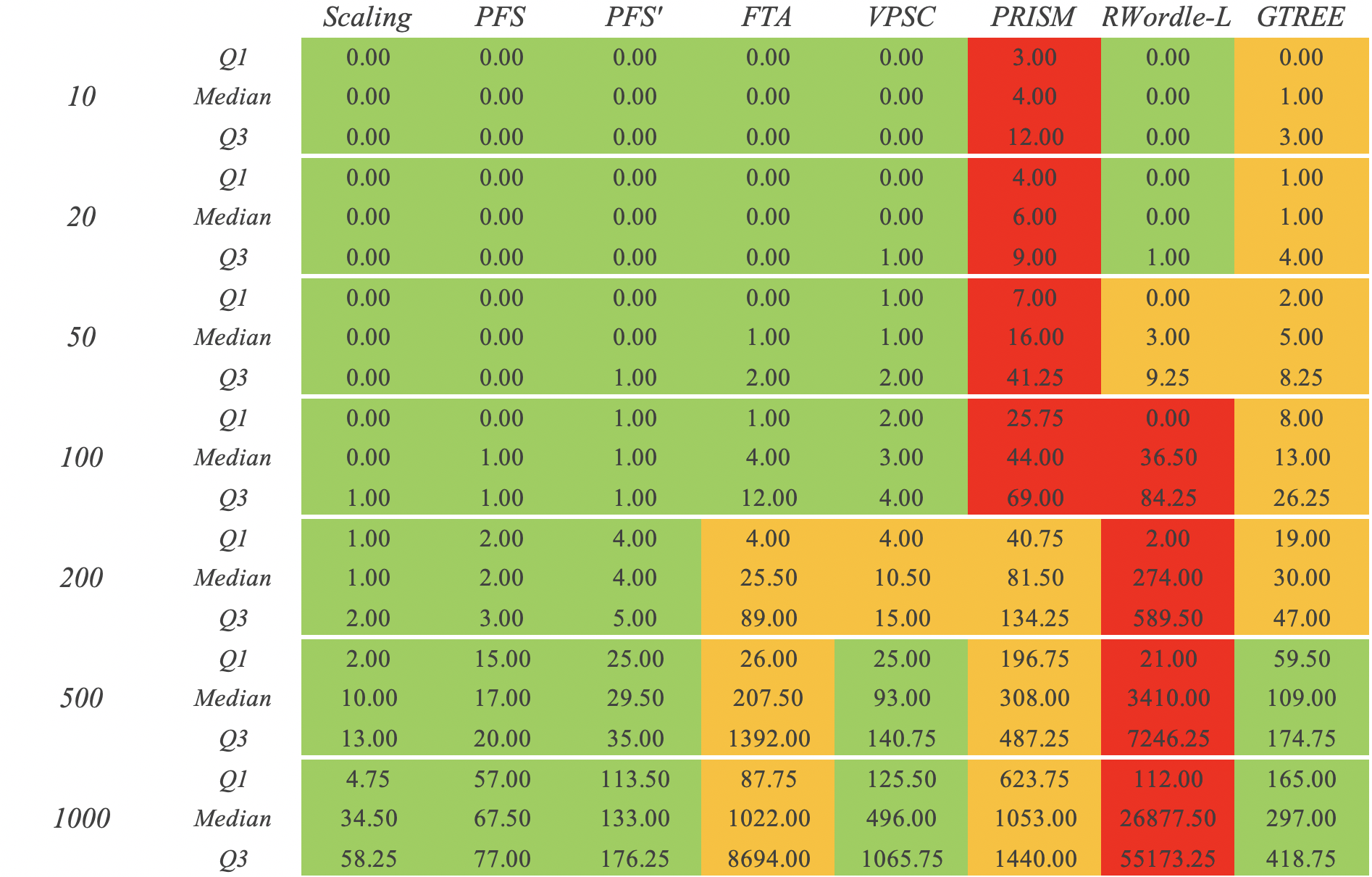}
    \caption{Aggregated running times among the synthetic graphs, function of number of nodes (10 to 1,000): first quartile, median and third quartile.}
    \label{fig:res_synt_time}
\end{figure}

\begin{figure}[ht!]
    \centering
    \includegraphics[width=0.9\linewidth]{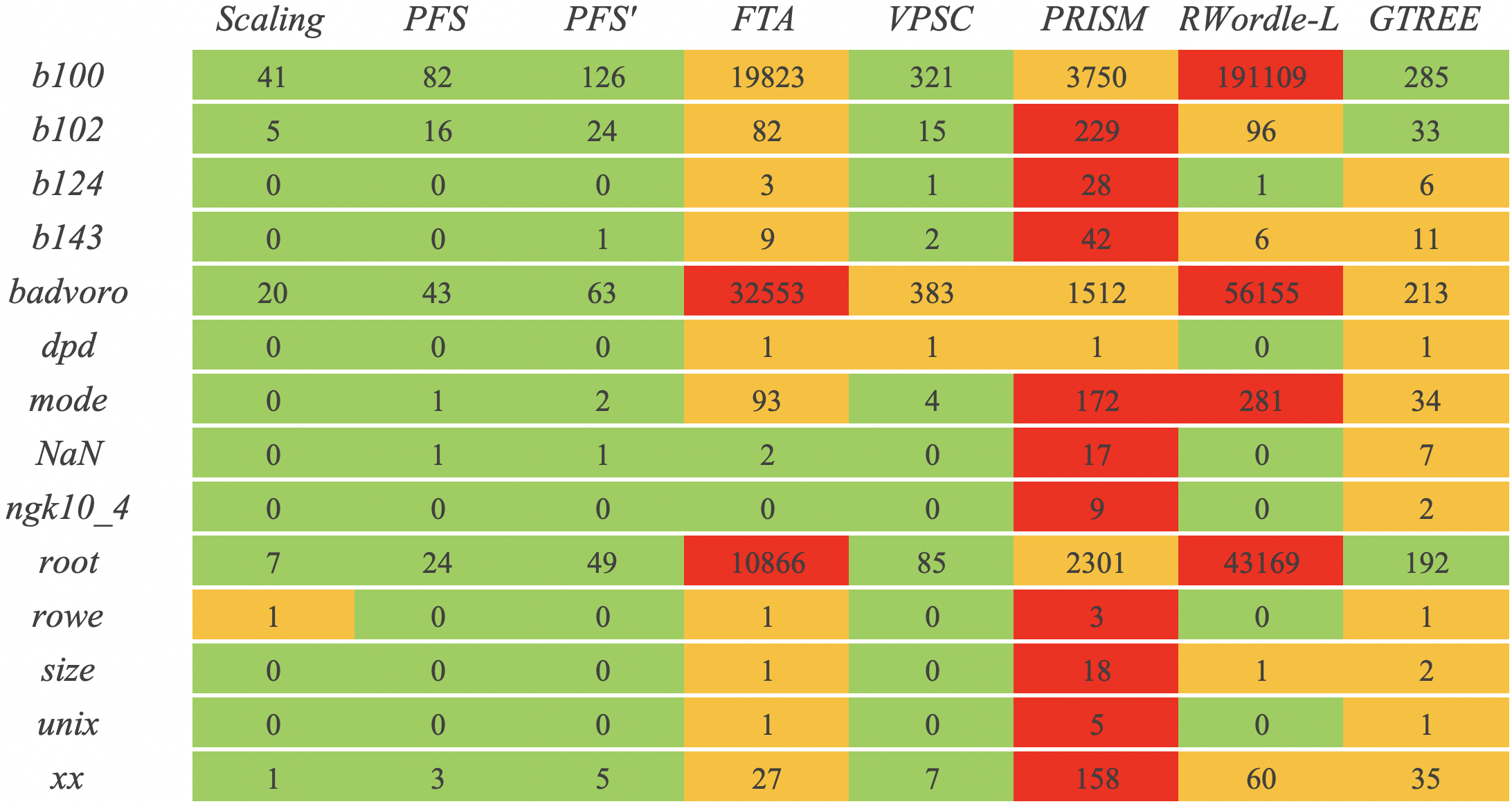}
    \caption{Mean values of running times among the real-world graphs.}
    \label{fig:res_rw_time}
\end{figure}

Fig.~\ref{fig:res_synt_time} and~\ref{fig:res_rw_time} show the aggregated running time values on the synthetic and real-world datasets. \textit{Scaling}, \textit{PFS}, \textit{PFS'} and \textit{VPSC} require lower running time. \textit{FTA} and \textit{GTREE} induce intermediate running time, but the third quartile shows that \textit{FTA} can induce a certain amount of time consuming embedding computations. Finally, \textit{PRISM} is time consuming for small graphs, but have intermediate results for larger graphs, while \textit{RWordle-L} has good results for small graphs but is very time-consuming for larger ones.
\section{Conclusion}\label{sec:conclusion}

As a conclusion, even if \textit{Scaling} optimises 4 out of 5 criteria and is very fast to compute on the graphs of our datasets, it does not represent a satisfying solution as it increases the size of the embedding too much. \textit{PFS} is also not satisfying as it got poor results on 3 criteria. \textit{FTA} obtained intermediate results over all the criteria, which is less good than all its remaining competitors. \textit{PFS'} and \textit{PRISM} obtained comparable results but the latter is more time-consuming. Both have intermediate results for shape preservation and node movement minimisation, which might be considered as two essential criteria. \textit{GTREE} suffers from inducing high node movements on our datasets. Overall, \textit{VPSC} and \textit{RWordle-L} obtained the best quality results. While \textit{RWordle-L} outperforms \textit{VPSC} on global shape preservation and is comparable on the other criteria, \textit{VPSC} outperforms \textit{RWordle-L} in terms of running time. Finally, considering the different types of graphs (random graphs, random trees, small world graphs, and scale-free graphs), we did not observe any significant differences in terms of results.

\subsubsection*{Acknowledgement} This research has been partly funded by a national French grant (ANR Daphne 17-CE28-0013-01). 

\bibliographystyle{splncs04}
\bibliography{bibliography}

\end{document}